\documentclass[aps,prl,twocolumn,showpacs,amssymb,amsmath]{revtex4}
\usepackage[dvips]{graphicx,color}
\usepackage{color,verbatim}

\begin{document}

\title{Sub-Poissonian phononic population in a
nanoelectromechanical system}

\author{Matteo Merlo$^{1,2}$, Federica Haupt$^{1}$, Fabio Cavaliere$^{1}$, and Maura Sassetti$^{1}$}

\affiliation{$^{1}$ Dipartimento di Fisica,
Universit\`a di Genova and  LAMIA INFM-CNR, Via Dodecaneso 33, 16146
Genova, Italy\\
 $^{2}$ Dipartimento di Fisica,
Universit\`a di Genova and INFN, Via Dodecaneso 33, 16146
Genova, Italy}

\begin{abstract}
Population of a phononic mode coupled to a single-electron transistor in the sequential tunneling regime
is discussed for the experimentally realistic case of intermediate electron-phonon coupling. Features
like a sub-Poissonian bosonic distribution are found in regimes where electron transport drives
the oscillator strongly out of equilibrium with only few phonon states selectively populated.
The electron Fano factor is compared to fluctuations in the phonon distribution,
showing that all possible combinations of sub- and
super-Poissonian character can be realized.
 \end{abstract}

 \date{\today}
\maketitle

%%%%%%%%%%%%%%%%%%%%%%%%%%%%%%%%%%%%%%%%%%%%%%%%%%%%%%%%%%%%%%%%%%%%%%%%
\noindent
%%%%%%%%%%%%%% INTRODUCTION %%%%%%%%%%%

\emph{Introduction.}--- Condensed-matter physics and optics have often exchanged concepts and ideas,
based on the common underlying structure of wave phenomena. These
 are essentially based on interference effects which, in the case of
light, can be derived from classical wave equations. However,
features exist which cannot be explained within a classical
treatment, as e.g. squeezed states~\cite{Breitenbach} and photon antibunching in
resonance fluorescence~\cite{Kimble}.  On the other hand, in mesoscopic physics the
focus is often on electron transport. The latter is naturally characterized in terms
of current, and much information (e.g. carriers charge, process statistics,
correlation effects) can be extracted from noise~\cite{Blanter}. Of particular
interest are then condensed-matter systems in which fermionic and bosonic degrees
of freedom are coupled, and where electron transport induces the emission of
nonclassical radiation. For example, it is known that electronic shot noise in
a quantum point contact   may be source of antibunched photons~\cite{Beenakker},
and  emission of antibunched phonons from a two-level quantum dot is expected
when transport is characterized by bunching of tunneling electrons~\cite{Brandes}.

Nanodevices where mechanical motion is coupled to electric transport constitute in this sense a
perfect subject for study~\cite{BlencoweRev}.  Realizations of these nanoelectromechanical systems (NEMS)
have been obtained e.g. with single oscillating  molecules~\cite{Park},
semiconductor beams~\cite{Knobel} and suspended carbon nanotubes~\cite{Leroy}.
NEMS are interesting dynamical systems and are expected to show many peculiar transport
features ranging from shuttling instability~\cite{shuttling} to avalanche-like
transport~\cite{Koch}. Recently,  attention has also been focused on the \emph{mechanical}
 properties of NEMS, as it appears now that experiments are close enough to the quantum
 limit~\cite{LaHaye} to test theoretically predicted quantum features in the vibrational
  motion~\cite{Ruskov}. Of particular interest are those associated with the discrete energy
  states of the oscillator and, indeed, several proposals have been put forward to measure
   discrete number states~\cite{Santamore&co}. Furthermore, it is well known that the distribution
   of oscillation quanta (phonons) in NEMS is strongly affected by the transport
   of electrons~\cite{Mitra,Lambert}, and even the existence of nonclassical number
    states induced by tunneling has been predicted~\cite{Rodriguez}.

In this Letter, we address the behavior of a harmonic oscillator coupled to a quantum dot,
focusing on the distribution of unequilibrated phonons
induced by electric transport. We show that it is possible to achieve a
\emph{selective} population of few phonon states such that the distribution of the phonon number $l$
displays a sub-Poissonian behavior, i.e. $\mbox{Var}\, l <\langle l \rangle$.
At the same time, we consider  the zero-frequency current noise  and show that the fluctuations
of both the phonon distribution and the electron current can be enhanced or reduced
with respect to Poissonian statistics one independently of the other.

%%%%%%%%%%%%%%%%%%%%%%%%%%%%%%%%%%%%%%%%%%%%%%%
%%%%%%%%%%%% MODEL AND METHODS%%%%%%%%%%%%%%%%%
%%%%%%%%%%%%%%%%%%%%%%%%%%%%%%%%%%%%%%%%%%%%%%%

\emph{Model \& methods.}---The system we consider is a gated single-electron transistor (SET)
coupled to leads and to a harmonic oscillator.
The SET Hamiltonian is described within the
standard constant-interaction model
 for spinless particles. In particular, the charging-energy term is
$H_{\it c}=E_{\it c}(n-n_{\it g})^2$, where $(n-n_{\it g})$ is  the effective number
of electrons on the SET and $n_{\it g}$ is proportional to the charge induced by the gate.
The oscillator and coupling terms are ($\hbar=1$)
\begin{eqnarray}
\label{H_S}
H_{\rm ph}=\omega_{0}\, b^\dagger b+
\sqrt{\lambda}\,\omega_{0}\,(b^{\dagger}+b)\,(n-n_{\it g}),
\end{eqnarray}
where $b^{\dagger}$ creates vibrational excitations of energy $\omega_{0}$.
The dimensionless parameter $\sqrt{\lambda}$ defines the strength of electromechanical interaction between the
position of the oscillator and the effective charge on the SET. Such a term can be induced
by an  oscillating gate capacitively coupled to the dot~\cite{Knobel}.
The leads are Fermi liquids with  $H_{\it l}=\sum_{k,\alpha=L,R}\varepsilon_{k,\alpha}c^{\dagger}_{k,\alpha
} c^{}_{k,\alpha}$ and chemical potentials $\mu_{L,R}=\mu_0\pm eV/2$, where
$V$  is the  bias voltage. In the limit $\omega_0, eV,k_B T \ll E_{\it c},\Delta E$,
 where $\Delta E$ is the average single-particle level spacing, the SET excess occupancy
 is limited to $0,1$ and we can focus  on the lowest unoccupied
single-particle level $\xi$. The total Hamiltonian can then be written as
$H= \epsilon n +H_{\rm ph}+H_{\it l}+H_{\it t}$, where  $\epsilon=\xi+2 E_{\it c}(1/2-n_{\it g})$
 and $n=d^{\dag}d$ are respectively the energy and the occupation number of the single
  level, and $H_{\it t}=\sum_{k,\alpha=L,R}(t_{\alpha}c^{\dag}_{k,\alpha}d+h.c.)$.
  Here, $t_{\alpha}$ are the tunneling amplitudes, with asymmetry $A=|t_{R}|^2/|t_{L}|^2$.

Being interested in the weak-tunneling limit, we  treat $H_{\it t}$ as a perturbation.
 It is then convenient to perform a canonical transformation to make the unperturbed Hamiltonian
 diagonal in the system variables $n,l$. The desired transformation is the Lang-Firsov polaron
  transformation $\bar{O}=UOU^{\dag}$ with $U=\exp{\eta(b-b^{\dag})}$ and $\eta =\sqrt{\lambda}
   (n-n_{\it g})$~\cite{Mitra}. The transformed Hamiltonian is given by
   $\bar{H}=\bar{\epsilon}n+\omega_0 b^{\dag}b+H_{\it l}+\bar{H}_{\it t}$, where $\bar{\epsilon}=\epsilon-\lambda \omega_0$ and
\begin{equation}
\bar{H}_{\it t}=\sum_{k,\alpha=L,R}(t_{\alpha}c_{k,\alpha}^{\dag}d\,e^{-\sqrt{\lambda} (b^{\dag}-b)} +h.c.).
\end{equation}
In the polaron picture, coherences between states with different phonon number
can be neglected as far as the level broadening $\gamma$ induced by tunneling
is the smallest energy scale into play~\cite{Mitra}, i.e. $\gamma \ll \omega_0,k_B T$.
 In this limit, the reduced density matrix $\bar{\rho}$ of the SET+oscillator system in
 the polaron picture is diagonal both in $n$ and $l$, and the dynamics is well described
 by the  rate equations
 $\partial_t \bar{P}_{0(1),l}=\sum_{l'}\big[\Gamma_{\it o\,(i)}^{l'l}\bar{P}_{1(0),l'}-\Gamma_{\it i\,(o)}^{ll'}\bar{P}_{0(1),l}\big]$
for the populations $\bar{P}_{n,l}= \langle n,l|\bar{\rho}|n,l\rangle$.
Here,  $\Gamma_{\it i\,(o)}^{ll'}= \sum_{\alpha}{\Gamma_{\!\!\alpha}}_{\it i\,(o)}^{ll'}$
 are the total rates for tunneling in (out of) the level, and
\begin{equation}
\begin{split}\label{rates}
{\Gamma_{\!\!\alpha}}_{\it i}^{ll'}=& \,\, 2 \pi \nu |t_{\alpha}|^2 X^{l'l}f_{\alpha}(\omega_0(l'-l))\\
{\Gamma_{\!\!\alpha}}_{\it o}^{ll'}=& \,\, 2\pi \nu |t_{\alpha}|^2 X^{l'l}[1-f_{\alpha}(\omega_0(l-l'))],
\end{split}
\end{equation}
where $f_{\alpha}(x)= f(x+\bar{\epsilon}-\mu_{\alpha})$ and  $f(x)$ is the Fermi function.
The coefficients $X^{ll'}= |\langle l' | e^{-\sqrt{\lambda} (b^{\dag}-b)}|l \rangle|^2$
are the Franck-Condon (FC) factors~\cite{Mitra,Koch} and $\nu$ is the density of states of
the leads.  In the following we assume  $\mu_0=\xi-\lambda \omega_0$ so that $n_{\it g}=1/2$ defines on-resonance conditions.
We focus on the regimes of  weak ($\lambda \ll 1$) and intermediate ($\lambda\approx 1$)
phonon coupling, where cotunneling is negligible  out of the Coulomb-blockaded regions~\cite{Koch}.

Electronic and phononic expectation values can be evaluated in the polaron picture as
$\langle O\rangle={\rm Tr}[\bar{O}\bar{\rho}]=\sum_{nl} \langle n,l |\bar{O}|n,l\rangle \bar{P}_{n,l}$.
 It is useful to define also a ``hybrid'' average  $\langle O\rangle_{\bar{\rho}}= {\rm Tr}[O\bar{\rho}]$.
 In terms of $\langle \cdot \rangle_{\bar{\rho}}$, we can write
\begin{eqnarray}
\langle l \rangle&=& \langle l \rangle_{\bar{\rho}}+\langle \eta^2 \rangle_{\bar{\rho}}, \label{lmedio} \\
\langle l^2 \rangle &=& \langle l^2 \rangle_{\bar{\rho}}+ 4\langle \eta^2 l \rangle_{\bar{\rho}} +
\langle \eta^2 \rangle_{\bar{\rho}}+\langle \eta^4 \rangle_{\bar{\rho}}, \label{l2medio}
\end{eqnarray}
where we have used the fact that $\bar{b}=b-\eta$ and that $\bar{\rho}$ is diagonal in the
 considered weak-tunneling limit. Note that for operators like $n$, which are unchanged by the canonical
 transformation, it is $\langle n\rangle_{\bar{\rho}}=\langle n\rangle$.

From the stationary solution of the master equation, the phonon Fano factor $F_{\rm ph}={\rm Var}\,l /\langle l \rangle$
 can be directly calculated in terms of  Eqs.(\ref{lmedio}) and (\ref{l2medio}).
The electronic Fano factor  $F= S/2 e \langle I\rangle$ is evaluated following Ref.~\onlinecite{Korotkov}.
\begin{figure}
\includegraphics*[bb=24 6 3036 1602,width=0.48\textwidth]{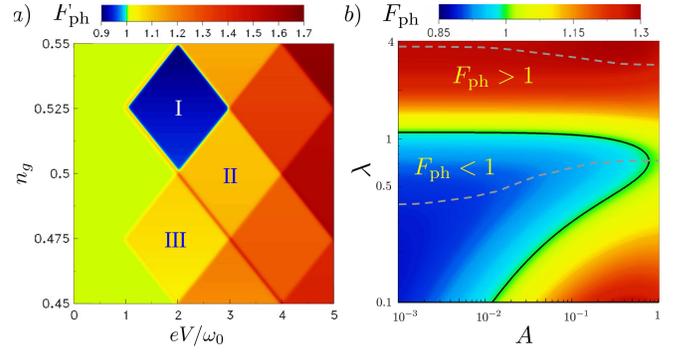}
\caption{{ $a$)} Phonon Fano factor $F_{\rm ph}$ as a function of
voltage $V$ and $n_{\it g}$, at $A=0.1$ and $\lambda=0.7$.
{ $b$)} Density plot of $F_{\rm ph}$ as a function of $A$ and $\lambda$ for $eV=2 \omega_0$, $n_{\it g}=0.525$ (middle of region I in $a$).
 Black line: contour $F_{\rm ph}=1$.
The region between the dashed lines encloses the four--state regime (see text). In both panels:
$k_B T=0.01\, \omega_0$, $E_{\it c}=10\,\omega_0$. Color scales on the top.
}\label{fani}
\end{figure}

%%%%%%%%%%%%%%%%%%%%%%%%%%%%
%%%%% Results %%%%%%%%%%%%%%
%%%%%%%%%%%%%%%%%%%%%%%%%%%%

\emph{Results.}--- We first consider the phonon Fano factor $F_{\rm ph}$. Our main result is
that, in the presence of asymmetry of the tunneling barriers and specific voltage conditions,
the phonon distribution shows a sub-Poissonian behavior $F_{\rm ph}<1$ (see Fig.~\ref{fani}$a$).

In particular, for $A<1$ ($A>1$) the most favorable region of the $V-n_{\it g}$ plane for having
$F_{\rm ph}<1$ is  region I (region III). A sub-Poissonian $F_{\rm ph}$ can also be obtained in
region II, but only in the limit of very strong asymmetry (not shown). For definiteness, in the
following we assume $A<1$ and we focus mainly on region I which, in general, is limited by the
following conditions: $\omega_0 \le eV \le 3\omega_0$ and $1/2 \le n_{\it g} \le 1/2+\omega_0/2 E_{\it c}$.

Here, the phonon Fano factor shows a crossover between sub- and super-Poissonian behavior as a
function of $A$ and $\lambda$ (see Fig.~\ref{fani}$b$). As a rule of thumb, a sub-Poissonian
$F_{\rm ph}$ requires $\lambda \lesssim 1$ and it is favored by strong asymmetries. Interestingly,
for intermediate values of the electron-phonon coupling $\lambda \approx 1$, it is $F_{\rm ph} <1$
 already for asymmetries which are experimentally feasible, $A\lesssim 1$.

Theoretically, the super- (sub-) Poissonian character of $F_{\rm ph}$ is more easily studied in
terms of the parameter $Q = {\rm Var}\,l-\langle l\rangle$, being $F_{\rm ph}<1$ only if $Q<0$.
Let us write $Q=Q_{\bar{\rho}}+\Delta Q$, where
$Q_{\bar{\rho}}=\langle l^2\rangle_{\bar{\rho}}-\langle l \rangle_{\bar{\rho}}^2 -\langle l \rangle_{\bar{\rho}}$ and
\begin{eqnarray} \label{deltaQ}
\Delta Q&=&2 \lambda \langle l \rangle_{\bar{\rho}} [n_{\it g}^2-(2 n_{\it g}-1)(2-\langle n\rangle_{\bar{\rho}})] \\
&+&4\lambda(2n_{\it g}-1)\langle l(1-n)\rangle_{\bar{\rho}} + \lambda^2(2 n_{\it g}-1)^2 {\rm Var}\,n. \nonumber
\end{eqnarray}
Taking into account that $n \in\{0,1\}$, it is easy to show that it is always $\Delta Q>0$ in region I.
 Therefore, a sub-Poissonian $F_{\rm ph}$ requires necessarily $Q_{\bar{\rho}}<0$.

The quantity $Q_{\bar{\rho}}$ can be  evaluated in terms of the phonon distribution in the polaron
frame $\bar{\mathcal{P}}_l=\bar{P}_{0,l}+\bar{P}_{1,l}$.  Note that if $\bar{\mathcal{P}}_l$ obeyed a thermal
distribution $\bar{\mathcal{P}}_l=e^{-\beta l\omega_0}(1-e^{-\beta \omega_0})$,
it would be $Q_{\bar{\rho}}=(e^{\beta \omega_0}-1)^{-2}\ge 0$, while if $\bar{\mathcal{P}}_l$ follows the
Poisson distribution it would obviously be $Q_{\bar{\rho}}=0$. On the other hand, when only the two lowest
vibrational levels are occupied,  $\bar{\mathcal{P}}_0+\bar{\mathcal{P}}_1=1$, it is $Q_{\bar{\rho}}=-\langle l\rangle_{\bar{\rho}}^2 \le 0$.

A phonon distribution having only the first few states occupied is therefore a promising candidate
for observing $F_{\rm ph}<1$. Indeed, our numerical investigations strongly suggest that a sub-Poissonian
 phonon Fano factor requires a slender phonon distribution: in fact  $F_{\rm ph}<1$ is solely observed
 at low voltages, in the presence of asymmetry and preferably off-resonance, which are all conditions
 which favor a narrow $\bar{\mathcal{P}}_l$~\cite{Mitra}.

\begin{figure}
\includegraphics*[width=0.42\textwidth]{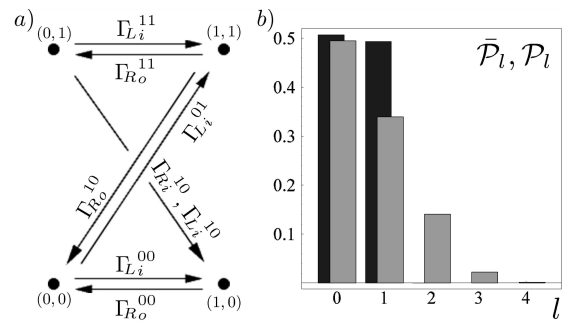}%\hspace{15mm}
\caption{{$a$)} Set of states included in the four--state model. States are labeled as $(n,l)$,
arrows represent the relevant transitions  in region I for $k_BT \ll \omega_0$.
{$b$)} $\bar{\mathcal{P}}_{l}$ (black) vs. $\mathcal{P}_l$ (gray) for $eV=2\, \omega_0$, $n_{\it g}=0.525$, $k_B T=0.01 \omega_0$,
$E_{\it c}=10\,\omega_0$ and $\lambda=0.9$, $A=0.01$.
}
\label{condition}
\end{figure}

Remarkably, for $\lambda=1$ the exact stationary solution of the rate equation satisfies the
condition $\bar{\mathcal{P}}_0+\bar{\mathcal{P}}_1=1$ always in region I.
This fact can be understood observing that in region I the only energetically allowed transitions
which increase the phonon number are $(0,l)\to (1,l+1)$, see Eq.(\ref{rates}).
As a consequence, at low temperatures $k_BT \ll \omega_0$,  excited phonons states can only be
populated  via a series of subsequent tunneling events such as $(0,0)\to(1,1)\to(0,1)\to(1,2)$.
However, for $\lambda=1$  the transition $(1,1)\to(0,1)$ is forbidden because the FC factor
 $X^{11}=e^{-\lambda}(1-\lambda)^2$ vanishes. In this case, the dynamics of the system is frozen
 to the states with $l\le1$ and it can be solved exactly by considering the reduced four--state model
  represented in Fig.~\ref{condition}$a$. Within this model, we obtain
\begin{equation} \nonumber
F_{\rm ph}\big|_{\lambda=1}=1+\frac{3 - 8 n_{\it g} + 4 n_{\it g}^2 + 2 A(3 - 8 n_{\it g} + 5 n_{\it g}^2)}{(2+A)[n_{\it g}^2(A+2)-4n_{\it g}+3]},
\end{equation}
which gives e.g.  $F_{\rm ph}=0.96$ for $n_{\it g}=0.54$ and $A=0.1$.

The four--state model of Fig.~\ref{condition}$a$ is often a good approximation of the full
 numerical solution also for $\lambda \neq 1$. Within this model, we derive an analytical expression
 for $F_{\rm ph}$ which, however, is too long to be reported here. This approximates the exact
 numerical result with an error smaller than 1\% in all region enclosed between the dashed
 lines in Fig.~\ref{fani}$b$, and therefore it allows us to investigate analytically the crossover
  between super- and sub-Poissonian phonon Fano factor. In particular, we find that it can be $F_{\rm ph}<1$ only
  for values of the electron-phonon coupling smaller than a certain critical value
   $\lambda_{\rm cr}$ which, up to order $(n_{\it g}-1/2)^2$, is given by
\begin{equation}
\lambda_{\rm cr}=\frac{1}{4(n_{\it g}-1)^2+2A(n_{\it g}^2-4n_{\it g}+2)}.
\end{equation}
This equation describes very accurately the upper part of the contour line $F_{\rm ph}=1$ in the
 phase diagram Fig.~\ref{fani}$b$,  deviating from the exact result only for  $A \to1$.

The existence of a maximum critical value $\lambda_{\rm cr}<1$ can be understood qualitatively
considering the limit of strong asymmetry $A\ll 1$. In this case, it is $\langle n \rangle \approx 1$ and $\mbox{Var}\, n\approx 0$,  and
from Eq.(\ref{deltaQ}) one obtains directly $Q\approx \langle l\rangle_{\bar{\rho}}[2 \lambda (n_{\it g}-1)^2-\langle l\rangle_{\bar{\rho}}]$,
 where we have used $Q_{\bar{\rho}}=-\langle l \rangle_{\bar{\rho}}^2$ in the four--state model.
A strong electron-phonon coupling is thus unfavorable for $Q<0$ in two respects: on one hand,
it increases the  weight of the positive term $\propto (n_{\it g}-1)^2$; on the other one, it
is well known that $\bar{\mathcal{P}}_{l=0}\to 1$ as $\lambda$ is increased~\cite{Mitra}, so that $\langle l \rangle_{\bar{\rho}}\to 0$.

We can conclude that the sub-Poissonian Fano factor is induced by a phonon distribution $\bar{\mathcal{P}}_l$
 in which only the first few phonon states are populated and yet the occupation probability of the
  excited states is comparable with the one of the ground state.  We refer to such a situation as a \emph{selective population} of the phonon states.

Finally, we remind that $\bar{\mathcal{P}}_l$ is the phonon distribution in the polaron picture.
 The intrinsic phonon distribution is $\mathcal{P}_l=\sum_n P_{n,l}$ where $P_{n,l}= \langle n,l |\rho |n,l\rangle$
 and $\rho=U^{\dag}\bar{\rho} \,U$ is the density matrix in the original picture.
Note that, in terms of $\mathcal{P}_l$, the average phonon number reads  $\langle l \rangle=\sum_l l \mathcal{P}_l$,
and  similarly $\langle l^2 \rangle =\sum_l l^2 \mathcal{P}_l$. We can evaluate $\mathcal{P}_l$ taking
 into account that $P_{n,l}=\sum_{l'}X_{n}^{ll'}\bar{P}_{n,l'}$, where $X_{n}^{ll'}=|\langle n,l |U |n,l' \rangle |^2$
  are generalized FC-factors. Such a relationship is a consequence of $\bar{\rho}$ being diagonal in the weak-tunneling limit.
Comparing  $\mathcal{P}_l$ and $\bar{\mathcal{P}}_l$, it is clear that one can speak of selective
population \emph{only} in the polaron picture (see Fig.~\ref{condition}$b$). However, what is important
is that, in the presence of a selective population of $\bar{\mathcal{P}}_l$,  the intrinsic phonon
distribution $\mathcal{P}_l$  shows a sub-Poissonian behavior, signaled by $F_{\rm ph}<1$.

%%%%%%%%%%%%%%%%%%%%%%%%%%%%%%%%%%
%%% Electronic Fano Factor %%%%%%%
%%%%%%%%%%%%%%%%%%%%%%%%%%%%%%%%%%
Up to now, we have   considered solely the characteristics of the phonon distribution induced by tunneling.
However, it is well known that the transport properties of the system are in turn strongly affected by phonons.
 Signatures of this interplay are especially visible in the current Fano factor, which is very
sensitive to the electron-phonon interaction~\cite{Koch,Haupt}. For example, a giant enhancement of $F$
 has been predicted as fingerprint of strong electron-phonon coupling~\cite{Koch}.
\begin{figure}[htbp]
\includegraphics*[bb=19 13 3015 1484, width=0.48\textwidth]{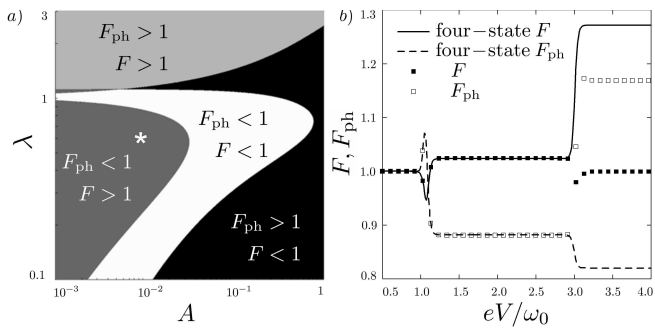}
\caption{{ $a$)} Phase diagram of the possible combinations of $F,F_{\rm ph}$  in region I ($eV=2\omega_0,n_{\it g}=0.525$)
depending on $\lambda$ and $A$ (see text for discussion).
{ $b$)} Plots of $F_{\rm ph}$ and $F$ as a function of $V$ for $n_{\it g}=0.525$
 and $\lambda=0.6$, $A=0.01$ corresponding to the asterisk in panel $a$ for $eV=2\omega_0$.
Boxes: exact numerical solutions; lines: four--state approximation.
In both panels $k_B T=0.01\,  \omega_0$, $E_{\it c}=10\,\omega_0$.}\label{fanicut}
\end{figure}
Here, we consider intermediate coupling %($\lambda$ around 1)
and we focus on the study of the (sub-) super-Poissonian character of $F$ with respect to the one of $F_{\rm ph}$.
  Interestingly, in region I  all the possible combinations of $F,F_{\rm ph} \lessgtr 1$ can be obtained
   by tuning the asymmetry $A$ and $\lambda$ (see Fig.~\ref{fanicut}$a$).
This is possible because the super- and sub-Poissonian character of $F$ and $F_{\rm ph}$ have different
 physical origins. In fact, while $F_{\rm ph}<1$ presumes a selective population of phonon states, $F>1$
  is induced by a bunching of tunneling events~\cite{Belzig}.

A simple explanation of this mechanism can be given in terms of the four--state model of Fig.~\ref{condition}$a$.
 For $\lambda \approx 1$, the electronic Fano factor can be written as:
\begin{equation} \nonumber
F=\frac{4+A^2}{(2+A)^2}+ \frac{4 + 2\,A + 14\,A^2 + 9\,A^3 - A^4}{{\left( 1 + A \right) }^2\,{\left( 2 + A \right) }^3}(1-\lambda)^2.
\end{equation}
For $\lambda=1$ the system behaves as a spin degenerate single level so that it is always $F \le 1$~\cite{Struben}.
 For $\lambda \approx 1$, the transitions $(0,0)\leftrightarrow(1,1)$ and $(0,0)\leftrightarrow(1,0)$ act
 as two competing transport channels, whose relative weight is determined by the ratio $X^{01}/X^{00}=\lambda$.
 It follows that for $\lambda<1$  the state $(1,1)$ is a trap state  and blocks the transport through
 the other more conducting channel $(0,0)\leftrightarrow(1,0)$. In the presence of asymmetry, such a
 dynamical channel blockade~\cite{Belzig} leads to bunching of tunneling events and to super-Poissonian
  current noise $F>1$. However, as the difference between the two competing transport channels is
   fairly weak, $F$ is only slightly above 1 (see Fig.~\ref{fanicut}$b$). The same mechanism occurs
    for $\lambda>1$ but, in this case, it is transport through the excited state that is blocked by the occupation of $(1,0)$.

It is then clear why super-Poissonian current noise and sub-Poissonian phonon distribution can
 occur simultaneously only for $\lambda<1$ when the trapping mechanism responsible for $F>1$ also
 favors the selective population of the phonon states. Viceversa, for $\lambda>1$ the occupation of
 the vibrational ground state is strongly favored since $(1,0)$ is the trap state, and  the
  phonon distribution is mainly super-Poissonian.
Note that outside region I the four--state model differs considerably from the exact results,
which exhibit $F\le 1$ and $F_{\rm ph}>1$ as  expected from a fermionic and bosonic system, respectively.

Finally, a comment is in order. A suppressed phonon Fano factor $F_{\rm ph}<1$ has recently been
 predicted for an oscillator driven by a superconducting SET in the limit $\gamma\sim \omega_0$,
  and this has been interpreted as signature of a number-squeezed state~\cite{Rodriguez}. In our case,
  instead,  we obtain a sub-Poissonian distribution \emph{without} squeezing. This is ultimately a
  consequence of the loss of phase information in the weak-tunneling limit $\gamma \ll \omega_0 $~\cite{Walls}.

%%%%%%%%%%%%%%%%%%%%%%%%%%%%
%%%% Conclusions %%%%%%%%%%
%%%%%%%%%%%%%%%%%%%%%%%%%%%%

In conclusion, we have shown that a sub-Poissonian phonon distribution can be achieved in
a nanoelectromechanical system when tunneling induces a selective population of few phonon states.
In addition, we have considered the electronic noise and we have found different combinations
of sub- and super-Poissonian electron and phonon Fano factors depending on the asymmetry and on the strength of the electron-phonon coupling,

Financial support by the EU via Contact No. MCRTN-CT2003-504574 and by the Italian MIUR via PRIN05 is gratefully acknowledged.

%%%%%%%%%%%%%%%%%%%%%%%%%%%%%%%
%%%% Bibliography %%%%%%%%%%%%%
%%%%%%%%%%%%%%%%%%%%%%%%%%%%%%%

\end{document}